\documentclass[epjST]{svjour}
\usepackage{graphics}
\usepackage{lineno}
  \usepackage{mathptmx}
\usepackage{subfigure}
\usepackage{dcolumn}
\usepackage{amsmath,amssymb}
\usepackage{bm}
\usepackage{color}
\usepackage{overpic}
\usepackage{latexsym}
\usepackage{epstopdf}
\usepackage{color}
\usepackage[english]{babel}
\usepackage{latexsym}
\usepackage{stmaryrd}

\usepackage{braket}

\definecolor{mygrey}{gray}{0.35}
\definecolor{myblue}{rgb}{0.2,0.2,0.8}
\definecolor{myzard}{cmyk}{0,0,0.05,0}
\definecolor{mywhite}{rgb}{1,1,1}
\definecolor{myred}{rgb}{1,0.,0.3}

\usepackage[colorlinks=true,citecolor=myblue,linkcolor=myred]{hyperref}

 \def\ee{\mathord{\rm e}}
 
 \def\ii{\mathord{\rm i}}
\def\min{\mathord{\rm min}}

\def\half{\textstyle\frac{1}{2}}

\renewcommand{\ii}{{\rm i}}
\renewcommand{\ee}{{\rm e}}
\begin{document}
\title{The Interspersed Spin Boson Lattice Model}

\author{A. Kurcz\inst{1}\and J.J. Garc\'ia-Ripoll\inst{1} \and A. Bermudez\inst{1}  }
%
\institute{Instituto de F\'isica Fundamental, IFF-CSIC, Calle Serrano 113b, Madrid E-28006, Spain}
\abstract{We describe a family of lattice models that  support a new class of  quantum magnetism characterized by correlated spin and bosonic ordering [\href{http://journals.aps.org/prl/abstract/10.1103/PhysRevLett.112.180405}{Phys. Rev. Lett. {\bf 112,} 180405 (2014)}]. We explore the full phase diagram of the model using  Matrix-Product-State methods. Guided by these numerical results, we describe a modified variational ansatz to improve our analytic description of the groundstate at low boson frequencies. Additionally, we  introduce an experimental protocol capable of inferring  the low-energy excitations of the system by means of Fano scattering spectroscopy. Finally, we discuss the implementation and characterization of this model with current circuit-QED technology.}

\maketitle
\setcounter{tocdepth}{2}
\begingroup
\hypersetup{linkcolor=myblue}
\tableofcontents
\endgroup

\section{Introduction}
\label{intro}

In the quest of the quantum computer, the development of quantum-information architectures  based on ultracold atoms~\cite{QS_ultracold_atoms}, trapped ions~\cite{QS_trapped_ions}, superconducting circuits~\cite{QS_cirquit_QED}, or arrays of quantum dots~\cite{QS_quantum_dots}, has reached  a point (or will probably do in a near future) where it is possible to isolate a quantum  system composed of many particles and control its dynamics such that it behaves according to a desired quantum many-body model~\cite{feynman}. These {\it quantum many-body platforms} have certain appealing features, such as the possibility of {\it (i)} performing experiments in a clean environment, {\it (ii)} designing the microscopic parameters of the model Hamiltonians,  {\it (iii)} initialising  and probing the system even at the single-particle level, and {\it (iv)} tracking its full dynamical evolution. This constitutes a unique oportunity where experimental results may guide the development of new analytical and numerical tools, and serve as a testbed for new theoretical ideas  addressing longstanding problems in quantum many-body physics~\cite{QS}. Possible examples are the relation of high-temperature superconductivity to the Hubbard model, the existence of thermalisation in closed many-body systems, or the interplay of interactions, frustration and quantum fluctuations in  quantum spin models.

In this article, we will focus on the latter. We note that, with the exception of \cite{ising_greiner}, the usual approach towards the realisation of magnetism in these quantum many-body platforms  relies on  perturbative processes that yield effective spin-spin interactions, such as  the super-exchange of atoms in optical lattices~\cite{superexchange_ol}, or  the  virtual exchange of phonons in ion crystals~\cite{Ising_ions}. In this case, one must work in a perturbative  regime in order to trace/integrate out the carriers and  obtain the desired spin  Hamiltonian~\cite{Rabi_dispersive,JC_dispersive} or Liouvillian~\cite{dissipative_xy}. This limits the strength of the spin-spin interactions,  posing thus  serious technological challenges to minimize other sources of technical or thermal noise, such that they lie  below  the  strength of spin-spin interactions. To overcome this limitation, one needs to abandon this perturbative regime, accepting thus that the carriers cannot be simply traced/integrated out. Instead, one is forced to consider them as part of the magnetic ordering of the system.   The main challenge is to   find  models where the bosons, rather than  destroying the  magnetic order, contribute to it yielding hybrid spin-boson magnets. In this work, we discuss certain models that lead to a spin-boson analogue of Ising magnets. We shall  focus on the {\it interspersed spin-boson (ISB) lattice model}~\cite{isblm_prl}, which we consider as a representative of these hybrid magnetic phases.

This paper is organised as follows. In Sec.~\ref{sec:2}, we introduce the ISB lattice model, and present analytical and numerical results that describe its full phase diagram. In Sec.~\ref{sec:3}, we discuss many-body spectroscopic protocols to probe the nature of these hybrid magnets. We turn into a possible implementation of the class of ISB lattice models in circuit-QED platforms in Sec.~\ref{sec:4}, and we present some conclusions and outlook in  Sec.~\ref{sec:5}.

\section{The Interspersed Spin Boson Lattice Model}
\label{sec:2}

\subsection{The model: definition and background}
\begin{figure*}
\centering
\includegraphics[width=0.8\columnwidth]{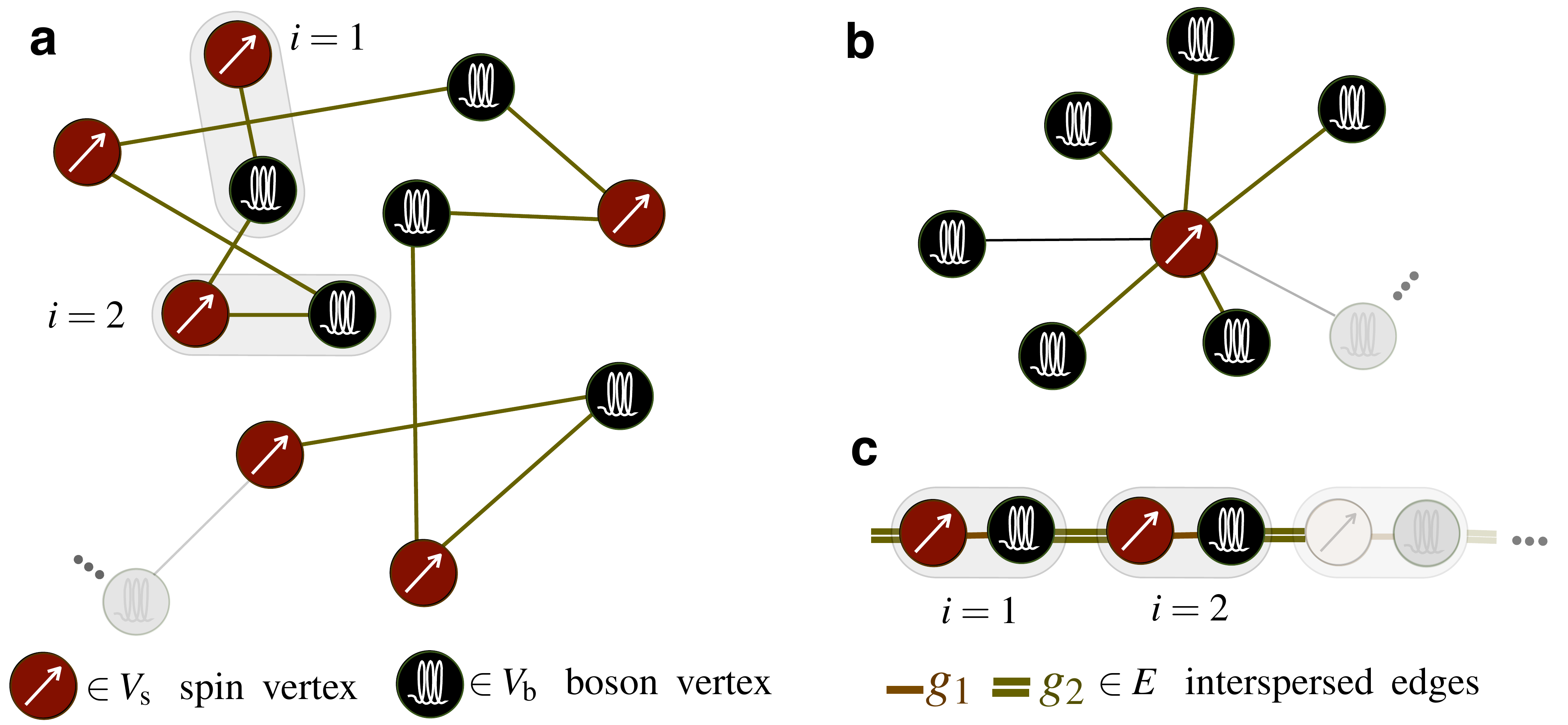}
\caption{ {\bf Scheme of the Interspersed Spin Boson Lattice model: } {\bf (a)} Scheme of the general ISB model~\eqref{eq:sblm} described by an undirected graph with spin and boson vertices joined by edges.  {\bf (b)} Spin-boson model for the dissipation of a two-level system coupled to a bosonic  environment. The  bosonic modes have different frequencies $\{\omega_b\}$, and  inhomogeneous couplings $\{g_b\}$ to the spin,  defining the bath spectral density $J(\omega)=\sum_b
|g_b|^2\delta(\omega-\omega_b)$. {\bf (c)}  Mapping of the ISB lattice model in {\bf (a)} to a 1D chain, where each unit cell contains a spin-boson dimer with nearest-neighbour spin-boson couplings.  }
\label{fig:isb_scheme}
\end{figure*}

Let us introduce the family of lattice models under consideration, which is defined by an undirected graph $G=(V, E)$, where the set of vertices $V=V_{\rm s}\bigcup V_{\rm b}$ results from the union two disjoint subsets $V_{\rm s}\bigcap V_{\rm b}=0$ (see Fig.~\ref{fig:isb_scheme}{\bf (a)}). The vertices  $s\in V_{\rm s}$ and $b \in V_{\rm b}$ host spin and bosonic degrees of freedom, respectively, which are  represented by  Pauli matrices  $\boldsymbol{\sigma}_s=({\sigma}^x_s,{\sigma}^y_s,{\sigma}^z_s)$ and  bosonic creation-annihilation operators $a^\dagger_b, a_b^{\phantom{^\dagger}}$. The set of edges $E$ is  composed by lines $e=(s,b)$ that connect  spin  $s\in V_{\rm s}$ and bosonic  $b\in V_{\rm b}$ vertices, defining thus the  spin-boson couplings. Finally, the dynamics of the model is dictated by the  spin-boson Hamiltonian
\begin{equation}
\label{eq:sblm_graph}
H_{\rm ISB}=\sum_{s\in V_{\rm s}}\frac{\omega_{s}}{2}\sigma_s^z+\sum_{b\in V_{\rm b}}\omega_b a^\dagger_b a_b^{\phantom{^\dagger}}+\sum_{(s,b)\in E} g_{s,b}\sigma_s^x(a_b^{\phantom{^\dagger}}+a_b^{\dagger}),
\end{equation}
where $\omega_s$ ($\omega_b$) is the spin (boson) resonance frequency, $g_{s,b}\in\mathbb{R}$ is the spin-boson coupling, and we set $\hbar=1$. Let us describe a particular example of Eq.~\eqref{eq:sblm_graph} that leads to the famous spin-boson model~\cite{sbm_legget}. This model,  which explores the dynamics of a quantum two-level system coupled  to an environment, can be obtained from  the  Hamiltonian~\eqref{eq:sblm_graph} by considering  a single spin coupled to a collection of bosons (i.e. each vertex $b\in V_{\rm b}$  corresponds to a different  mode of the environment, as in Fig.~\ref{fig:isb_scheme}{\bf (b)}). By including more spins, and shaping the  spin-boson couplings appropriately, we will go beyond this model, accessing  the physics of strongly-correlated systems and quantum phase transitions in hybrid spin-boson magnets.

To illustrate the richness of this family of interspersed spin boson  (ISB) lattice models~\eqref{eq:sblm_graph}, and to connect to existing literature, we focus on a graph where the spins are only connected to a pair of bosons and viceversa  (see  Fig.~\ref{fig:isb_scheme}{\bf (a)}). We can put this model on a chain $z_i=id$ with lattice spacing $d$, and  $i\in\{1\cdots N\}$, where each of the spins has two neighbouring bosons,  and the unit cell contains  a  spin-boson pair (see Fig.~\ref{fig:isb_scheme}{\bf (c)}). The ISB lattice model thus becomes
\begin{equation}
H_{\rm ISB}=\sum_{i}\frac{\omega_0}{2}\sigma_i^z+\sum_{i}\omega a^\dagger_i a_i^{\phantom{^\dagger}}+\sum_{i}\sigma_i^x( g_1a_i^{\phantom{^\dagger}}+g_2a_{i-1}^{\phantom{^\dagger}} +{\rm H.c.}),
\label{eq:sblm}
\end{equation}
where the spin (boson) resonance frequency $\omega_0$ ($\omega$) is homogeneous, and  the  sign of $g_1,g_2\in\mathbb{R}$  determines the ferro ($g_1/g_2>0$) or anti-ferro ($g_1/g_2<0$) spin-boson ordering. We set $|g_1|=|g_2|=:g$ without loss of generality, and focus on the anti-ferro regime $g_1=-g_2=g$. 

At this point, let us  compare our model to the  {Jaynes-Cummings} and {Rabi lattice models}, which  arise  in coupled-cavity arrays~\cite{JC_lattice_photons,Rabi_lattice_photons}, or  trapped-ion crystals~\cite{JC_lattice_ions,Rabi_lattice_ions}, and may be considered as particular instances of cooperative Jahn-Teller models~\cite{jahn_teller_models}. In these models,   spins only interact locally with the bosonic modes, which  are in turn coupled  among themselves.  In the limit of perturbative spin-boson couplings, the bosons act as mediators of long-range spin-spin interactions, either for the Jaynes-Cummings~\cite{JC_dispersive} or the Rabi~\cite{Rabi_dispersive} lattice. In contrast, bosons in the perturbative regime of the ISB lattice model~\eqref{eq:sblm} yield short-range nearest-neighbour interactions. This particular feature allowed us to develop an analytical theory beyond the perturbative regime (i.e. ultra-strong coupling $g\sim\omega,\omega_0$ )~\cite{isblm_prl} which, supplemented by extensive numerical simulations, showed that the ISB lattice model hosts a hybrid spin-boson quantum phase transition within the Ising universality class.  This phenomenon coincides with previous predictions  for the Rabi lattice model~\cite{Rabi_lattice_photons,Rabi_lattice_ions}, indicating that both systems should lie in the same universality class. Just as the nearest-neighbour quantum Ising model, due to its analytical tractability~\cite{pfeuty_ising},  enjoys a privileged position within a universality class that  hosts  models with long-range interactions~\cite{ising_long_range}; we consider that the ISB lattice model~\eqref{eq:sblm} is the prototype of a  class of hybrid spin-boson quantum magnets displaying a $\mathbb{Z}_2$ quantum phase transition, which also includes models with  longer range  such as~\cite{Rabi_lattice_photons,Rabi_lattice_ions}. 

In  section~\ref{sec:2.1} below, we will start with a brief review of our analytical theory for the ISB  groundstate~\cite{isblm_prl}, where we studied  ultrastrong couplings $|g|\lesssim\omega$, but limited ourselves to high bosonic frequencies $\omega_0<\omega$. Then, we shall explore numerically the remaining part of the phase diagram  $\omega_0>\omega$, and show that our variational ansatz still works considerably well except for $\omega\ll\omega_0$. In that particular regime, we modify our ansatz and show that the  phenomenology of a spin-boson quantum magnet still applies. This complements  our previous work~\cite{isblm_prl}, and constitutes a full study of the phase diagram of the ISB lattice model.
Finally, in section~\ref{sec:2.2}, we discuss the analytical theory for the low-energy excitations.

\subsection{Variational ansatz for the many-body groundstate}
\label{sec:2.1}
\subsubsection{Lang-Firsov-type ansatz}
We use a type of Lang-Firsov transformation,  first introduced in the context of polarons in electron-phonon systems~\cite{lang_firsov}, by adapting it to the ISB model on the chain~\eqref{eq:sblm}. This unitary transformation, which can be expressed as 
\begin{equation}
 U_{\rm LF}= \ee^{-\ii \sum_i \frac{\Theta_i}{2}\sigma_i^x},\hspace{2ex} \Theta_i= -\ii\frac{2 g}{\omega}\left( a_i^{\phantom{^\dagger}}-a_{i-1}^{\phantom{^\dagger}}- a^\dagger_i+a^\dagger_{i-1}\right),
 \label{eq:LF}
 \end{equation}
   allows us to define a family of variational states $ \ket{\Psi_{\rm GS}^{\rm LF}} = U^{\dagger}_{\rm LF} \ket{\psi_{\rm spin}} \bigotimes_i\ee^{\alpha_i a_i^\dagger-\alpha_i^* a_i^{\phantom{^\dagger}}}\ket{0_i},$  where $\ket{0}_i$ is the vacuum of the corresponding boson. This ansatz is composed of a  spin state $ \ket{\psi_{\rm spin}}$ with an exponentially-large number of  complex variational parameters, and a product of bosonic coherent states with $N$ real variational parameters. Due to the Lang-Firsov unitary 
 \begin{equation}
 U_{\rm LF} a_i^{\phantom{^\dagger}} U_{\rm LF}^{\dagger}= a_i^{\phantom{^\dagger}}-\frac{g}{\omega}(\sigma_i^x-\sigma_{i+1}^x),\hspace{2ex}U_{\rm LF} \sigma_i^z U_{\rm LF}^{\dagger}=\cos\Theta_i\sigma_i^z+\sin\Theta_i\sigma_i^z,
 \label{eq:LF_transf}
 \end{equation}
and considering that  $\bra{\alpha_i}\cos\Theta_i\ket{\alpha_i}=\ee^{-4(g/\omega)^2}$ and $\bra{\alpha_i}\sin\Theta_i\ket{\alpha_i}=0$ for $\alpha_i\in\mathbb{R}$,  the variational minimisation $E_{\rm GS}^{\rm LF}=\min\{\bra{\Psi_{\rm GS}^{\rm LF}} H_{\rm ISB}\ket{\Psi_{\rm GS}^{\rm LF}} \}$ leads the bosonic vacuum $\alpha_i=0$ in the transformed picture. Additionally,  the spin state corresponds to the groundstate of an antiferromagnetic  Ising model with an exponentially-renormalised transverse field (rTIM), namely
\begin{equation}
 H_{\rm rTIM} = J\sum_{i}  \sigma^x_i\sigma^x_{i+1}
 +h _{\rm t} \sum_i \sigma^z_i ,\hspace{3ex} J=\frac{2g^2}{\omega},\hspace{1 ex}h_{\rm t}=\frac{\omega_0}{2}\ee^{-\frac{4g^2}{\omega^2}}.
 \label{H_rQIM}
\end{equation}
As advanced in the introduction, we obtain only nearest-neighbour interactions, which  contrasts to the long-range spin-spin interactions that appear in the Jaynes-Cummings~\cite{JC_dispersive} and Rabi~\cite{Rabi_dispersive} lattice models. This feature allows for a complete analytical solution of the variational ansatz~\cite{isblm_prl} after  the standard Jordan-Wigner~\cite{jordan_wigner} and Bogoliubov~\cite{bogoliubov} transformations. Let us leave the explicit expressions for the Jordan-Wigner-Bogoliubov fermions to the section on the low-energy excitations, and focus instead  on the prediction of the critical line and the hybrid nature of the spin-boson Ising magnet. Our ansatz predicts a correlated anti-ferromagnetic ordering of both the spin and the boson 'polarisations'
\begin{equation}
\langle\sigma_i^x\rangle_{\rm LF}=(-1)^i\left(1-\lambda^2\right)^{1/8}\theta(1-\lambda),\hspace{3ex} \langle a_i^{\phantom{^\dagger}}\rangle_{\rm LF}=(-1)^{i+1}\frac{2g}{\omega}\left(1-\lambda^2\right)^{1/8}\theta(1-\lambda),
\label{LF_polarizations}
\end{equation}
where $\lambda={h}_{\rm t}/J$, and $\theta(x)$ is the Heaviside step function that  yields a critical line at ${h}_{\rm t}=J$. Let us emphasise that, as announced in the introduction, the bosons of the  ISB lattice model are not merely carriers to be traced/integrated out, but instead  contribute to the hybrid spin-boson anti-ferromagnetic ordering when ${h}<J$. For the ISB model, such an order corresponds to the N\'{e}el alternation of the direction of both the spin projection and bosonic displacement.

\begin{figure*}
\centering
\includegraphics[width=0.8\columnwidth]{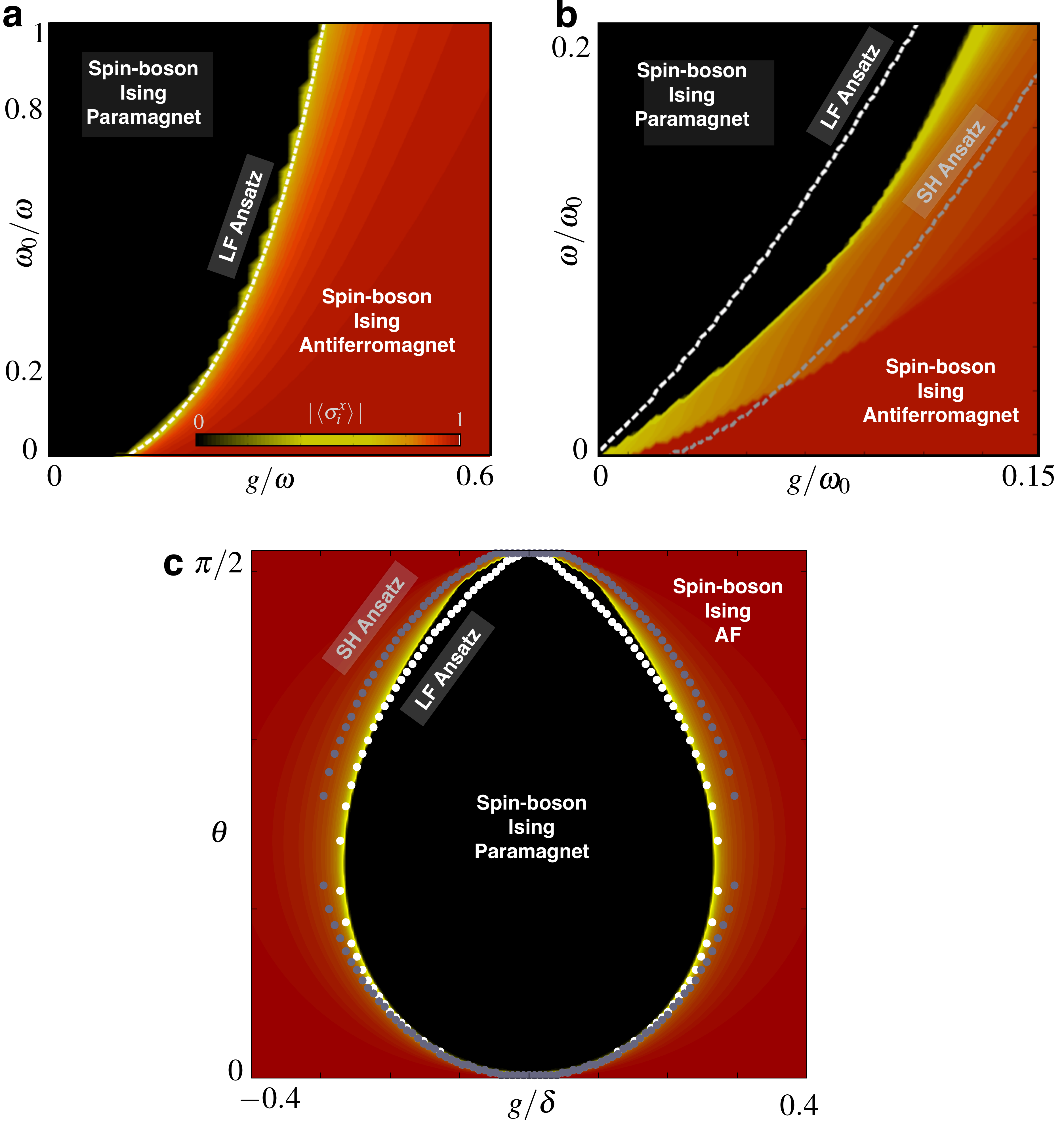}
\caption{ {\bf Phase diagram of the Spin Boson Lattice model: } {\bf (a)} Magnetization $\langle \sigma_i^{x}\rangle $ of the ISB groundstate for large bosonic frequencies $\omega>\omega_0$. The critical line predicted by Eq.~\eqref{LF_polarizations} corresponds to the white dashed line. {\bf (b)}  Small bosonic frequencies. {\bf (c)} Full phase diagram of the ISB lattice model with the parameterization   $\omega=\delta\cos\theta$, and $\omega_0=\delta\sin\theta$.}
\label{fig:isb_phase_diagram}
\end{figure*}

  In Fig.~\ref{fig:isb_phase_diagram}, we benchmark our ansatz  with  the numerical results obtained with a routine based on Matrix-Product-States, the so-called iTEBD~\cite{itebd,garcia_ripoll_2006,verstraete_2008}. In the regime $\omega>\omega_0$ studied in~\cite{isblm_prl}, the agreement with our ansatz is almost perfect  (Fig.~\ref{fig:isb_phase_diagram}{\bf (a)}). However, we have now explored  the regime $\omega<\omega_0$, where  some discrepancies start arising  in a small region around $\omega\ll\omega_0$ (Fig.~\ref{fig:isb_phase_diagram}{\bf (b)}). This is not surprising in light of the infrared divergence of the Lang-Firsov displacements~\eqref{eq:LF_transf} and the vanishing of the transverse field~\eqref{H_rQIM} as $\omega\to 0$, an effect which has already been identified in the standard spin boson model~\cite{sbm_legget}, and underlies the total breakdown of the Lang-Firsov ansatz. In contrast to this model, where the spin couples to a finite density of low-frequency bosons,  our spins  only couple to a pair of low-frequency modes. Moreover,  the spin-boson coupling is a free parameter in our model (i.e. not fixed by a certain spectral density $J(\omega)\propto\omega^{s}$ as in~\cite{sbm_legget}). Hence, it is always possible to lower $g$ in order to attain the critical region $J\sim h_{\rm t}$ even for $\omega\ll\omega_0$. We believe that it is the combination of these two properties  which reduces the spurious effects of the infrared  divergence, such that the Lang-Firsov ansatz does not break down completely. Anyhow, one could try to improve the Lang-Firsov ansatz  even further by applying techniques developed in the context of the spin-boson model~\cite{sbm_legget} to overcome the infrared divergence. This is the goal of the following section.
 
\subsubsection{Silbey-Harris-type ansatz}
Paralleling the previous section, we introduce a type of Silbey-Harris transformation,  introduced in the context of Kondo~\cite{luther_emery} and  spin-boson~\cite{silbey_harris} models. This variational transformation can be adapted to  the ISB model~\eqref{eq:sblm} as follows
\begin{equation}
 U_{\rm SH}(f)= \ee^{-\ii \sum_i \frac{\Theta_i}{2}\sigma_i^x},\hspace{2ex} \Theta_i= -\ii\frac{2 f}{\omega}\left( a_i^{\phantom{^\dagger}}-a_{i-1}^{\phantom{^\dagger}}- a^\dagger_i+a^\dagger_{i-1}\right),
 \label{eq:SH}
 \end{equation}
 where $f$  is no longer fixed, but  a variational parameter. In the  spin-boson model~\cite{silbey_harris}, the variational minimisation yields  a different displacement $f(\omega)$  for each bosonic mode. For large boson frequencies $f\to g$, and one recovers the Lang-Firsov displacement. On the contrary, the displacement  vanishes $f\to 0$ for small frequencies, since  the low-frequency bosons cannot adapt adiabatically to the spin. It is this behaviour  which cures the infrared divergence.  
 
  In addition to the Silbey-Harris unitary, we use a spin-boson inversion for  even  sites
 \begin{equation}
 U_{ {\rm e}\mathbb{Z}_2}=\ee^{-\ii \frac{\pi}{2}\sum_i\sigma^z_{2i}}\ee^{\ii\pi\sum_i a_{2i}^\dagger a_{2i}^{\phantom{^\dagger}}}.
 \end{equation}
This last transformation encodes the anti-ferromagnetic alternation,   allowing us to consider translationally-invariant  coherent states  $ \ket{\Psi_{\rm GS}^{\rm SH}} = U_{{\rm e}\mathbb{Z}_2}^{\dagger}U^{\dagger}_{\rm SH} (f)\ket{\psi_{\rm spin}} \bigotimes_i\ee^{\alpha a_i^\dagger-\alpha^* a_i^{\phantom{^\dagger}}}\ket{0_i}$. In contrast to the previous Lang-Firsov ansatz, this new ansatz has an additional variational parameter $f$, and a  single  real variational parameter for the bosonic coherent states. Using the  counterpart of the transformations~\eqref{eq:LF_transf}, the variational minimisation $E_{\rm GS}^{\rm SH}=\min\{\bra{\Psi_{\rm GS}^{\rm SH}} H_{\rm ISB}\ket{\Psi_{\rm GS}^{\rm SH}} \}$ yields now a nearest-neighbour Ising model in   mixed transverse and longitudinal  fields
 \begin{equation}
 H_{\rm rTLIM} = J(f)N+\omega\alpha^2N+J(f)\sum_{i}  \sigma^x_i\sigma^x_{i+1}
 + {h}_{\rm t}(f) \sum_i \sigma^z_i+{h}_{\ell}(f,\alpha) \sum_i \sigma^x_i ,
  \label{H_rTLIM}
\end{equation}
where the different constants now depend on the variational parameters
\begin{equation}
J(f)=\frac{2f(f-2g)}{\omega},\hspace{3 ex}{h}_{\rm t}(f)=\frac{\omega_0}{2}\ee^{-\frac{4f^2}{\omega^2}}, \hspace{3 ex}{h}_{\ell }(f,\alpha)=4\alpha(g-f).
\end{equation}

Unfortunately, this model is no longer exactly solvable via the Jordan-Wigner mapping~\cite{pfeuty_ising}. The so-called string operator that appears when expressing the longitudinal field $\sigma_i^x$ in terms of fermions $c_i^{\phantom{^\dagger}},c_i^\dagger$, namely $\sigma_i^x=\otimes_{j<i}(1-2c_j^\dagger c_j^{\phantom{^\dagger}})(c_i^{\phantom{^\dagger}}+c_i^{\dagger})$ makes the problem non-quadratic. 
Therefore, in order to carry on further with the ansatz, we will have to make some  approximations. We use a sort of Hartree-Fock decoupling $\sigma_i^x\to \langle\sigma_i^x\rangle=\left(1-\lambda_{\rm t}^2\right)^{1/8}\theta(1-\lambda_{\rm t})$, where $\lambda_{\rm t}={h}_{\rm t}(f)/|J(f)|$, and the expectation value of the transverse magnetization is calculated  for the Ising model with vanishing longitudinal field.
As we will see from the the numerical benchmark, this rather drastic approximation turns out to capture the low-frequency physics of the problem slightly better than the previous ansatz. Altogether, the variational energy that must be minimised is
\begin{equation}
E_{\rm GS}^{\rm SH}(f,\alpha)=\left(J(f)+\omega\alpha^2 -\frac{2J(f)}{\pi}(1+\lambda_{\rm t})\mathcal{E}(\theta_{\rm t})+4\alpha(g-f)\left(1-\lambda_{\rm t}^2\right)^{1/8}\theta(1-\lambda_{\rm t})\right)N,
\end{equation}
where we have introduced the elliptic integral $\mathcal{E}(\theta_{\rm t})=\int_0^{\pi/2d}d q(1-\theta_{\rm t}^2\sin^2(qd))^{1/2}$ with $\theta_{\rm t}=(4\lambda_{\rm t}/(1+\lambda_{\rm t})^2)^{1/2}$. From the variational minimisation, we find  $(f^\star,\alpha^\star)$, and then recover the critical line by solving ${h}_{\rm t}(f^\star)=J(f^\star)$. 

The results displayed in Fig.~\ref{fig:isb_phase_diagram}{\bf (b)} show that, in the regime $\omega\ll\omega_0$, this new ansatz is somehow complementary to the previous Lang-Firsov theory. While this theory overestimates the amount of anti-ferromagnetic correlations, the Silbey-Harris ansatz underestimates it. In Fig.~\ref{fig:isb_phase_diagram}{\bf (c)}, we  display the full phase diagram obtained by parameterizing the resonance frequencies $\omega=\delta\cos\theta$, and $\omega_0=\delta\sin\theta$, in terms of a common strength $\delta$ and angle $\theta\in[0,\pi/2]$. We observe in this figure that around $\theta\sim \pi/2$, where the spin frequency dominates, the Silbey Harris ansatz performs slightly better than the Lang-Firsov. Probably, a more elaborate treatment of the longitudinal field in Eq.~\eqref{H_rTLIM} would lead to a better agreement of the ansatz in all parameter regimes. For instance, one could try an approach where $\alpha=2(g-f)\langle \sigma_i^x\rangle$ is substituted in the longitudinal field of Eq.~\eqref{H_rTLIM}, where  $\langle \sigma_i^x\rangle$ is obtained from the Ising model without longitudinal field. Then, by numerically solving  the Ising model with mixed fields, one could calculate  $\langle \sigma_i^x\rangle$  and use it to iterate the procedure, and find the self-consistent solution. However, the numerical step would require again Matrix-Product-State methods, and  we would loose the  analytical appeal of the variational ansatz. Finally, let us note that another  possibility, especially for more complicated graphs of the ISB model where the spins may be coupled to several bosonic modes, would be to enlarge the Silbey-Harris ansatz by introducing several displacements with different weights~\cite{multipolaron}. 

\subsection{Variational ansatz for the low-energy excitations}
\label{sec:2.2}

In this subsection, we show how to build a variational ansatz for the low-energy excitations~\cite{isblm_prl}.  Let us rewrite the  groundstate obtained from the variational minimisation, either based on Lang-Firsov or Silbey-Harris transformations, as  $ \ket{\Psi_{\rm GS}} = U^{\dagger}\ket{\Omega}$. Here,  $\ket{\Omega}$ will play the role of a reference state in the definition of the variational  low-energy excitations
\begin{equation}
 \ket{\Psi_{\rm exc}} = U^{\dagger}\sum_{q\in {\rm HBZ}}\bigg(\beta_{\rm f}(q)\gamma_{q+}^{\dagger}+\beta_{{\rm b}}(q)a_q^{\dagger}\bigg)\ket{\Omega},\hspace{2ex} {\rm HBZ}=\bigg[0,\frac{\pi}{d}\bigg).
 \label{eq:ansatz_exc}
\end{equation}
where we have introduced  variational parameters  $ \beta_{\rm b}(q), \beta_{\rm f}(q)$ for  bosonic and spin excitations, together with their corresponding creation operators  in quasi-momentum space
\begin{equation}
\begin{split}
 a_{q}^{\dagger}=\frac{1}{\sqrt{N}}\sum_j\ee^{\ii qdj}a_j^{\dagger},\hspace{2ex} \gamma_{q+}^{\dagger}=\frac{1}{\sqrt{N}}\sum_j\ee^{\ii qdj}({u}_qc^{\dagger}_j+{v}_q^*c_j^{\phantom{^\dagger}}),
 \end{split}
\end{equation}
where we have introduced the parameters listed in~\cite{comment_excitations}. We can now express the variational energy  $\mathcal{E}(\{\boldsymbol{\beta}_q^{\dagger},\boldsymbol{\beta}_q\})= \bra{\Psi_{\rm exc}} H_{\rm ISB} \ket{\Psi_{\rm exc}}=\sum_q\boldsymbol{\beta}_q^{\dagger} \mathbb{H}_q\boldsymbol{\beta}_q$ as a quadratic functional of $\boldsymbol{\beta}_q=(\beta_{\rm f}(q),\beta_{\rm b}(q))^t$, where  $\mathbb{H}_q$ is a  Hermitian matrix. The variational minimisation amounts to solving a simple eigenvalue problem $(\mathbb{H}_q-\epsilon_{\rm exc}(q)\mathbb{I})\boldsymbol{\beta}_q=0$ for each quasi-momentum.

For instance, for the Lang-Firsov ansatz~\cite{isblm_prl}, we find the low-energy excitations  by diagonalizing a two-band model $\mathbb{H}_q\ket{\epsilon^{\rm LF}_{{\rm exc},\pm}(q)}=(E_{\rm GS}^{\rm LF}+\epsilon^{\rm LF}_{{\rm exc},\pm}(q))\ket{\epsilon^{\rm LF}_{{\rm exc},\pm}(q)}$, namely 
\begin{equation}
\mathbb{H}_q=\left(\begin{array}{cc}\omega_q & {g}_q \\ {g}^*_q & \epsilon_q\end{array}\right),\hspace{2ex}\epsilon_{{\rm exc},\pm}^{\rm LF}(q)=\half{(\omega_q+{\epsilon}_q)\pm\half\sqrt{(\omega_q-{\epsilon}_q)^2+4|g_q|^2}}.
\label{anstaz_energies}
\end{equation}
Here, $\omega_q=\omega+4{h}_{\rm t}(2g/\omega)^2\sin^2 (qd/2)$ is the dispersion relation of a purely bosonic excitation, ${\epsilon}_q=2[(J\cos qd+{h}_{\rm t})^2+(J\sin qd)^2]^{1/2}$ corresponds to the dispersion of  a purely spin excitation, and ${g}_{q}={h}_{\rm t}(2g/\omega)(1-\ee^{-\ii qd})({u}_q+{v}_q^*)$ is a coupling between the boson- and spin-like excitations. As a consequence of this coupling, the boson and spin dispersions get mixed, such that the low-energy quasiparticles  carry simultaneously spin and bosonic 'polarisations'. 

We describe below two spectroscopic protocols that yield the excitation energies in terms of some dynamical observables, and can be exploited as: {\it (i)}  a numerical benchmark of the ansatz, or {\it (ii)} an experimental guide to probe the many-body properties of the ISB lattice model. In Sec.~\ref{sec:3.1}, we  will  review briefly the spectroscopy protocol based on linear response theory presented in~\cite{isblm_prl}. Such a protocol may be quite demanding in terms of experimental resources, as it requires time-resolved measurements of the bosons of every lattice site. We thus introduce in Sec.~\ref{sec:3.2} a  simpler spectroscopic protocol based on resonance scattering, which only requires the measurement of the a single transmission coefficient.

\section{Many-body Spectroscopic Protocols}
\label{sec:3}

\begin{figure*}
\centering
\includegraphics[width=0.65\columnwidth]{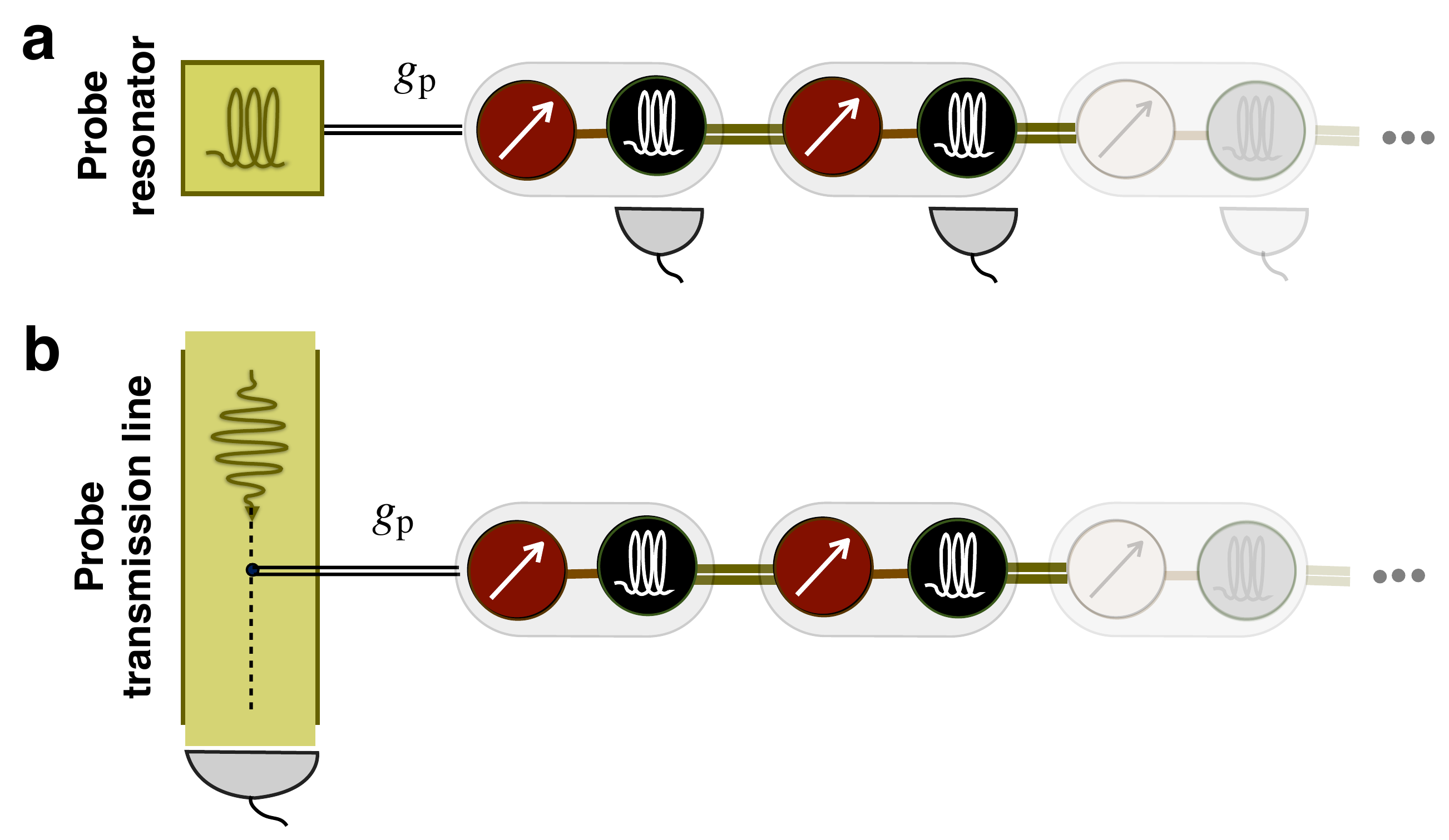}
\caption{ {\bf Many-body spectroscopy of  the interspersed spin boson lattice model: }  Scheme for the spectroscopy protocol based on Kubo linear response {\bf (a)} and  Fano resonant scattering  {\bf (b)}.  }
\label{fig:spectroscopy_scheme}
\end{figure*}

We  describe a general formalism  to extract the energies of the spin-boson excitations $\epsilon_{{\rm exc},\pm}(q)$, although the ideas presented herein can be easily generalised to other lattice models. Let us  introduce a  {\it non-invasive quantum probe}  that is: {\it (i) } described by a certain Hamiltonian $H_{\rm p}$, and {\it (ii)} coupled  perturbatively  to the first spin of the ISB lattice model $V_{\rm p-ISB}={g}_{\rm p}X_{\rm p}\sigma_1^x$. Here, $X_{\rm p}$ is a particular operator of the quantum probe, and the system-probe coupling ${g}_{\rm p}$ must be sufficiently weak  such that the spin-boson excitations do not get modified (see Fig.~\ref{fig:spectroscopy_scheme}). The composite system is then initialised in $\ket{\Psi(0)}=\ket{\Psi_{\rm p}}\otimes\ket{\Psi_{\rm GS}}$, where the groundstate of the ISB lattice model $\ket{\Psi_{\rm GS}}$ is prepared by adiabatic evolution to the desired parameter regime $(\omega,\omega_0, g)$, and $\ket{\Psi_{\rm p}}$ contains some excitation of the probe. This excitation will be transferred onto the system through the system-probe coupling $V$, thus inducing a dynamical response in the system.   Finally, we  measure an observable, either belonging to the system $O_{\rm ISB}$  (Sec.~\ref{sec:3.1}) or to the probe $O_{\rm p}$(Sec.~\ref{sec:3.2}), which must encode the desired  information about $\epsilon_{{\rm exc},\pm}(q)$.

\subsection{Kubo linear-response spectroscopy}
\label{sec:3.1}

We consider a   probe consisting of a single bosonic resonator $H_{\rm p}=\omega_{\rm p}a^\dagger_{\rm p}a_{\rm p}^{\phantom{^\dagger}}$, where $\omega_{\rm p}$ is the resonator frequency, and its coupling to the system is due to a Rabi-type interaction $X_{\rm p}=a_{\rm p}^{\phantom{^\dagger}}+a_{\rm p}^{\dagger}$ (Fig.~\ref{fig:spectroscopy_scheme}{\bf (a)}). The probe is initialised in a coherent state $\ket{\Psi_{\rm p}}={\rm exp}\{\alpha_{\rm p}(a_{\rm p}^{\dagger}-a_{\rm p})\}\ket{0_{\rm p}}$ with a low mean number of bosons warranted by setting $\alpha_{\rm p}\ll 1$, such that only a fewquasiparticles  will get excited in the system. Finally, we measure a system observable  built with $\{\langle {a}_j(t)\rangle\}$. The main idea is that, due to the system-probe perturbation, the change of the observable with respect to the unperturbed case shall encode the quasiparticle energies at first-order in the perturbation (i.e. linear response).

\begin{figure*}
\centering
\includegraphics[width=0.9\columnwidth]{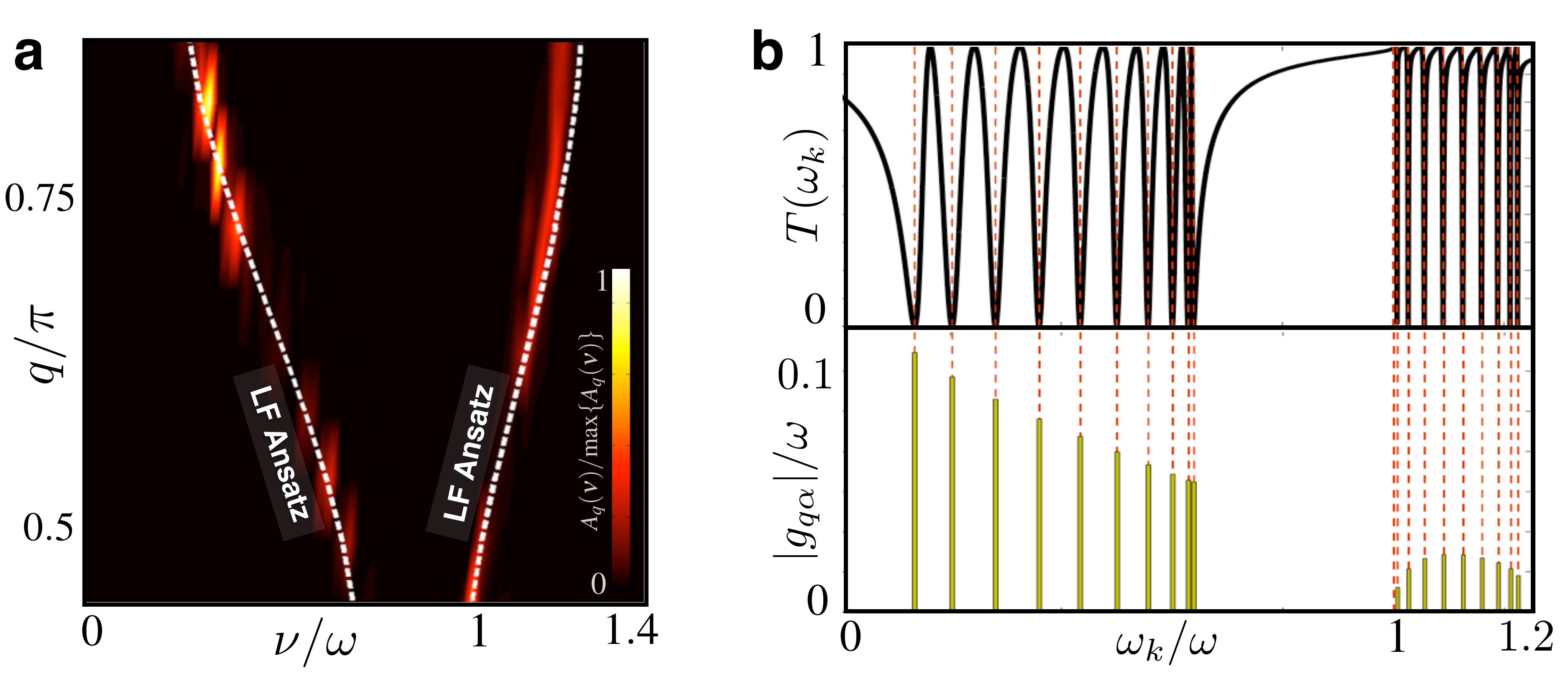}
\caption{ {\bf Many-body spectroscopic protocols: } {\bf (a)} Observable $\mathcal{A}_q(\nu)$ as a function of momentum and energy in the Kubo linear-response protocol for a ISB model with $N=10$ sites. The peaks serve to benchmark the ansatz energies in Eq.~\eqref{anstaz_energies}, which are represented as a dashed white line.  {\bf (b)}  Transmission coefficient $T(\omega_k)$ as a function of the incoming boson energy in the Fano resonant-scattering protocol for a ISB model with $N=10$ sites.  The dips in the upper pannel match perfectly the ansatz energies~\eqref{anstaz_energies}. In the lower pannel, we observe that $g_{k+}=0$ for the quasiparticle in the upper band with $k=0$. This is  why the corresponding dip in the transmission spectrum is absent.}
\label{fig:isb_spectroscopy}
\end{figure*}

The dynamical response of the system can be calculated along the same lines as Kubo's linear response theory~\cite{kubo}. A  Fourier transform $\mathcal{A}_k(\nu)=\int_0^{\infty}\ee^{\ii \nu t}\sum_j\ee^{-\ii k j}\langle a_j(t)\rangle/\sqrt{N}$  to  the momentum-frequency space  $(k,\nu)$ can indeed be expressed as 
\begin{equation}
\mathcal{A}_k(\nu)= \mathcal{A}_\pi\delta(k-\pi)\delta(\nu)+\ii\alpha_{\rm p}\delta(k-q)\left( G^{\rm b}_{q,{\rm p}}(\nu)+\chi_q G^{\rm f}_{q,{\rm p}}(\nu)\right),
\label{response_function}
\end{equation}
where we have introduced the system-probe retarded Green's functions~\cite{isblm_prl}, namely
\begin{equation}
G^{\rm b}_{q,{\rm p}}(\nu)=-\ii \int_{-\infty}^{\infty}\hspace{-1.5ex}{\rm d}t \ee^{\ii\nu t}\theta(t)\left\langle [a_q(t),a_{\rm p}^{\dagger}(0)]\right\rangle,\hspace{1.5ex} G^{\rm f}_{q,{\rm p}}(\nu)=-\ii \int_{-\infty}^{\infty}\hspace{-1.5ex}{\rm d}t \ee^{\ii\nu t}\theta(t)\left\langle [\tilde{\gamma}_{q,+}(t),a_{\rm p}^{\dagger}(0)]\right\rangle.
\label{sp_gf}
\end{equation}
These retarded Green's functions describe how an excitation, initially created at the probe resonator, travels through the system in the form of a boson- or spin-like excitation. 
They can be calculated analytically using our dynamical ansatz~\eqref{eq:ansatz_exc}, which shows that they have poles  at the quasiparticle energies above the groundstate~\eqref{anstaz_energies}, as usually occurs in other many-body systems~\cite{bruss_flensberg}. These poles translate into peaks of the response function~\eqref{response_function}, which get  broadened and shifted  due to the system-probe coupling. Therefore, a non-invasive quantum probe is obtained when $|g_{\rm p}|\ll \epsilon_{{\rm exc},\pm}(q)$, such that the level shift and broadening can be neglected. In Fig.~\ref{fig:isb_spectroscopy}{\bf (a)} , we display the response function~\eqref{response_function} obtained from a numerical simulation with matrix product states. The peaks obtained numerically are used to benchmark the validity of our dynamical ansatz~\eqref{eq:ansatz_exc}, and bring the possibility of measuring some critical exponents of the model to test if it lies in the Ising universality class.

As mentioned previously, this spectroscopy protocol requires a considerable amount of time- and single-site-resolved measurements, which might be quite demanding from an experimental perspective. In the following section, we will introduce a protocol that relaxes these requirements (i.e. no time-resolved measurements, and no single-site addressability), and still gives us access to the excitation energies, albeit without momentum resolution.

\subsection{Fano resonant-scattering spectroscopy}
\label{sec:3.2}

In this case, we consider a quantum probe that consists of a one-dimensional waveguide (e.g. open transmission line), where bosons travel with a well-defined group velocity $v_{\rm g}$ in  two possible directions. This is described by the following Hamiltonian
 \begin{equation}
 H_{\rm p}=\int {\rm d}x\left(\Phi_{\rm R}^{\dagger}(x)(\omega_{\rm p}-\ii \mathrm{v}_{\rm g}\partial_x))\Phi_{\rm R}(x)+\Phi_{\rm L}^{\dagger}(x)(\omega_{\rm p}+\ii \mathrm{v}_{\rm g}\partial_x))\Phi_{\rm L}(x)\right),
 \end{equation}
 where $\Phi_{\rm R}(x),\Phi_{\rm L}(x)$ are the quantum fields of right- and left-moving bosons, respectively, and $\omega_{\rm p}$ serves to set correctly their energy  (i.e. the above linear dispersion $\omega_k=\omega_{\rm p}-\mathrm{v}_{\rm g}k$ is typically an approximation of the waveguide spectrum around a certain frequency-momentum). The system-probe coupling $X_{\rm p}=\int {\rm d}x\delta(x-x_{\rm s})\left(\Phi_{\rm R}(x)+\Phi_{\rm L}(x)+{\rm H.c.}\right)$ is again due to a Rabi-type interaction  (see Fig.~\ref{fig:spectroscopy_scheme}{\bf (b)}), and we have set $x_{\rm s}$ as the point of the waveguide where the system is coupled. The probe is initialised as an incoming single-photon state with  well-defined momentum $\ket{\Psi_{\rm p}}=\int {\rm d}x \ee^{\ii k(x-x_{\rm s})}\Phi_{\rm R}^{\dagger}(x)\ket{0_{\rm p}}$, and we shall measure the transmission coefficient  far away from the system  $x_{\rm m}\gg x_{\rm s}$, namely $T(\omega_k)=\langle \Phi_{\rm R}^{\dagger}(x_{\rm m})\Phi_{\rm R}(x_{\rm m})\rangle$. The main idea of this scheme is that the incoming boson will scatter off the system when approaching a resonance with a quasiparticle excitation, and this will modify the transmission.  

This scattering setup is reminiscent of the so-called Fano resonance~\cite{fano,fano_resonances}, since the incoming boson has two possible paths  across the scattering point $x_{\rm s}$, either through the waveguide or after an excursion inside the system. There can be thus destructive interference leading to transmission and reflection with the so-called Fano line shapes. For a single resonant two-level scatterer, this effect can lead to a perfect mirror~\cite{shen_fang_ol,fano_anderson_scattering}. We shall determine below the conditions required to have a  resonance transmission probe of   the quasiparticle energies $\epsilon_{{\rm exc},\pm}(q)$ of the ISB lattice model~\eqref{eq:sblm} based on this effect. Let us also mention that,  if our ISB scatterer is connected to a pair of waveguides at both of its extremes in a sort of transport experiment~\cite{fano_metamaterial}, one would obtain a similar spectroscopic tool where the roles of the transmission and reflection are reversed when reaching a resonance.

Let us consider our dynamical ansatz~\eqref{eq:ansatz_exc}, and   focus on scattering close to the quasiparticle resonance  $\omega_k\approx\epsilon^{\rm LF}_{{\rm exc},\pm}(q)$. Moreover, by setting   $|g_{\rm p}|\ll \omega_k< \omega_k+\epsilon^{\rm LF}_{{\rm exc},\pm}(q)$, the system-probe coupling is not sufficiently energetic  to create higher-energy excitations~\cite{comment_high_energies}. We can then project the dynamics  onto the low-energy manifold $\mathcal{V}_\ell={\rm span}\{\ket{\Psi_{\rm GS}^{\rm LF}},\ket{\epsilon_{\pm}^{\rm LF}(q)}\}$, namely 
\begin{equation}
\begin{split}
H^{\ell}_{\rm ISB}&=\sum_{q}\sum_{\alpha=\pm}\int {\rm d}x\delta(x-x_{\rm s})\epsilon_{{\rm exc}, \alpha}^{\rm LF}(q)\sigma_{q,\alpha}^+\sigma_{q,\alpha}^-,\\
V^{\ell}_{\rm p-ISB}&=\sum_{q}\sum_{\alpha=\pm}\int {\rm d}x\delta(x-x_{\rm s})g_{q\alpha}\sigma_{q,\alpha}^+(\Phi_{\rm R}(x)+\Phi_{\rm L}(x))+{\rm H.c.},
\end{split}
\end{equation}
where  we have shifted the energy zero to $E_{\rm GS}^{\rm LF}$,  $\sigma_{q,\alpha}^+=\ket{\epsilon_{\alpha}^{\rm LF}(q)}\bra{\Psi_{\rm GS}^{\rm LF}}=(\sigma_{q,\alpha}^+)^\dagger$ are ladder operators that connect the groundstate to a low-energy excitation, and the coupling strengths are  $g_{q\alpha}$ are defined in~\cite{comment_coupling_fano}. Note that we have used the above constraints on the coupling strength to neglect the off-resonant terms that do not conserve the total number of excitations (i.e.  quasiparticles plus waveguide bosons).
 
Once this low-energy Hamiltonian has been found, one can solve the  Schr\"{o}dinger equation for a stationary scattering state, and then use the Lipmann-Schwinger formalism~\cite{shen_fang_pra} to connect it to the outgoing scattering state. In particular, the transmission coefficient far away from the scatterer is found to be $T(\omega_k)=|t(\omega_k)|^2$, where
\begin{equation}
t(\omega_k)=\left(1+\ii\sum_{q\alpha}\frac{\Gamma_{q\alpha}}{\omega_k-\epsilon_{{\rm exc}, \alpha}^{\rm LF}(q)}\right)^{-1},\hspace{2ex}\Gamma_{q\alpha}=\frac{|g_{q\alpha}|^2}{{\rm v_g}}.
\end{equation}
It is clear from this expression that, whenever the incoming boson hits a quasiparticle resonance $\omega_k\to\epsilon_{{\rm exc}, \alpha}^{\rm LF}(q)$, the transmission  $T(\omega_k)\to 0$ vanishes, and one has a Fano probe of the particular  energy (Fig.~\ref{fig:isb_spectroscopy}{\bf (b)}). To have well-resolved transmission dips, the broadening should fulfil $\Gamma_{q\alpha}\ll (\epsilon_{{\rm exc}, \alpha}^{\rm LF}(q)-\epsilon_{{\rm exc}, \alpha}^{\rm LF}(q'))$, which imposes  further constraints on  the  probe.

As announced previously, this new spectroscopic protocol based on Fano resonance scattering is less demanding from a experimental point of view, as it requires the measurement of a single probe observable at a single scattering time. On the other hand, it requires some tunability of the waveguide parameters $\omega_{\rm p},{\rm v_g},k$ where the incoming single photon is prepared.

\section{Implementation in circuit-QED}
\label{sec:4}

The interspersed spin-boson model admits a very natural implementation using superconducting circuits. These are centimeter size circuits of superconducting materials that work under cryogenic conditions ($T \sim 10-50$mK), so that thermal excitations neither break the superconducting pairs nor populate the typical circuit resonances ($\omega\sim 1-10$GHz). In such a setup, therefore, voltage and intensity or charge and flux constitute a pair of canonically quantized variables that can be rigorously studied using an effective quantum theory\ \cite{devoret95,yurke84} that allows for the design of microwave cavities, propagating microwave photons and also artificial two- or few-level systems or qubits.

Our interspersed model can thus be constructed in two different ways~\cite{comment_implementation}. A very simple one consists of superconducting striplines or coplanar resonators that are coupled to superconducting qubits, one cavity being shared by two qubits, and each qubit talking to two cavities. The first part of this setup, that is two qubits talking to the same cavity, has already been demonstrated\ \cite{majer07}, while the second part of this setup, that is two cavities talking to the same qubit, is routinely used in readout protocols for transmon qubits\ \cite{sun13}. Indeed, this brings us to another possible setup for our model, which would be three-dimensional superconducting cavities~\cite{3d_cavities}, where a macroscopic superconducting qubit sit in between cavities, mediating their interaction, much like in the chip. While this setup is not as mature as superconducting chips, it has the advantage that cavities and qubits have very long coherence times, approaching milliseconds, and allowing for very accurate writing, storage, readout and manipulation of all degrees of freedom.

\section{Conclusions and Outlook}
\label{sec:5}

We have described a new type of spin-boson magnetism that can be realised in trapped-ion or cirquit-QED quantum many-body platforms. We have introduced the family of interspersed spin boson lattice models~\eqref{eq:sblm_graph}, and  focused on the 1D chain as a testbed to develop a numeric and analytical toolbox to understand and probe such spin-boson magnets. However, we emphasise that this particular example does not exhaust all the richness of the ISB models~\eqref{eq:sblm_graph}, and we believe it would be interesting to explore other effects. For instance, introducing additional spin-boson couplings  in Fig.~\ref{fig:isb_scheme}{\bf (c)} can lead to a neat platform to study frustrated spin-boson magnets. The next step along this line would be to combine chains forming ladders with a variable number of rungs to finally reach the two-dimensional limit. A different avenue of research would be   to consider that each bosonic site (Fig.~\ref{fig:isb_scheme}{\bf (a)})  hosts a collection of bosons with a defined spectral density. This many-body generalisation of the spin-boson model would allow to study the interplay of  spin-boson magnetism and dissipation.


\begin{thebibliography}{}
\bibitem{comment_excitations}
We define the  following parameters ${u}_q=[\half(1+{\Delta}_q/{\epsilon}_q)]^{1/2}$, ${v}_q=\ii [\half(1-{\Delta}_q/{\epsilon}_q)]^{1/2}$ in terms of ${\Delta}_q=2(J\cos qd+{h}_{\rm t})$, and ${\epsilon}_q=[{\Delta}_q^2+(2J\sin qd)^2]^{1/2}$.
\bibitem{comment_high_energies}
The lowest higher-energy excitations are composed of two-quasiparticles with energies $\epsilon^{\rm LF}_{{\rm exc},\pm}(q,q')=\epsilon^{\rm LF}_{{\rm exc},\pm}(q)+\epsilon^{\rm LF}_{{\rm exc},\pm}(q')$. Since the system-probe coupling may connect a low-energy excitation~\eqref{eq:ansatz_exc} to these higher-energy ones, we have to impose that the process is highly off-resonant $|g_{\rm p}|\ll |\epsilon^{\rm LF}_{{\rm exc},\pm}(q,q')-\epsilon^{\rm LF}_{{\rm exc},\pm}(q'')|,\forall, q, q', q''$, which amounts to the condition in the text.
\bibitem{comment_coupling_fano}
The multi-level Fano  coupling constants are   $g_{q\alpha}=g_{\rm p}\langle e_2\ket{\epsilon_{\alpha}^{\rm LF}(q)}(u_q\delta_{\alpha,+}-v_q\delta_{\alpha,-})\ee^{\ii qd}/\sqrt{N}$, where $\bra{ e_2}=(0,1).$
\end{thebibliography}
\end{document}